\documentclass[11pt,twoside]{article}

\usepackage{asp2004} 
\usepackage{graphicx}
\usepackage{psfig}
\usepackage{epsf}
\usepackage{lscape}

\def\edcomment#1{\iffalse\marginpar{\raggedright\sl#1\/}\else\relax\fi}
\reversemarginpar

\newcommand{\Mpc}{$h^{-1}$\thinspace Mpc}

\def\apj{ApJ}
\def\apjl{ApJL}
\def\apjs{ApJS}

\def\aap{A\&A}
\def\mnras{MNRAS}

\begin{document}

\title{Clusters and Superclusters in the Sloan Survey}

\author{Jaan Einasto, Erik Tago, Maret Einasto, \& Enn Saar}
  \affil{Tartu Observatory, EE-61602 T\~oravere, Estonia}

\begin{abstract}
We find clusters and superclusters of galaxies using the Data Release
1 of the Sloan Digital Sky Survey. We calculate a low-resolution
density field with a smoothing length of 10~\Mpc\ to extract
superclusters of galaxies, and a high-resolution density field with a
smoothing length of 0.8~\Mpc\ to see the fine structure within
superclusters.  We found that clusters in a high-density environment
have luminosities that are about five times higher than the
luminosities of clusters in a low-density environment.  Numerical
simulations show that in large underdense regions most particles form
a rarefied population of pregalactic matter whereas in large overdense
regions most particles form a clustered population in rich clusters.
Simulations show also that very massive superclusters are great
attractors and have small bulk motions.  Less massive superclusters
are smaller attractors and have much larger bulk motions.
\end{abstract}

\thispagestyle{plain}


\section{Introduction}

Clusters and superclusters of galaxies are the basic building blocks
of the Universe on cosmological scales.  The first catalogues of
clusters of galaxies (Abell \cite{abell}, Zwicky et al. \cite{zwicky})
were constructed by visual inspection of the Palomar Observatory Sky
Survey plates.  Cluster catalogues have been used to define
superclusters of galaxies (Oort~\cite{oort83}, Bahcall~\cite{b88},
Einasto et al. \cite{e1994},~\cite{e1997}, ~\cite{e2001}, hereafter
E94, E97 and E01, Basilakos~\cite{bas03}).

In the present study we found catalogues of groups/clusters and
superclusters using galaxy samples of the Data Release 1 of the Sloan
Digital Sky Survey (DR1 of SDSS). For comparison we use the
cluster catalogue of the 2 degree Field (2dF) redshift survey by Eke
et al. \cite{eke04}. These data enable us to investigate the properties of
clusters of galaxies in various environment, from rich superclusters
to poor filaments of loose groups in voids.  For comparison we also
use clusters and superclusters found in N-body simulations of
evolution of structure. The present study is a continuation of the study
of clusters and superclusters based on the Early Data Release of SDSS and
the Las Campanas Redshift Survey by Einasto et al. \cite{e03a},
\cite{e03b}, \cite{e03c}, \cite{e03d} (hereafter E03a, E03b, E03c and
E03d, respectively), and Hein\"am\"aki et al. \cite{hei03}.  E03a and
E03b  found clusters of galaxies as density enhancements in the
high-resolution density field.  In the present study we shall define
groups and clusters in the conventional way using 3-dimensional data
on the distribution of galaxies.
The overall distribution of superclusters in space can be best studied 
using the Abell superclusters (E94, E97 and E01); the data used here
allow a more detailed investigation of the fine structure of
superclusters.   A Powerpoint version of the talk
with colored figures is available on the web-sites of Tartu Observatory
(http://www.aai.ee/$\sim$einasto) and of the conference
(http://mensa.ast.uct.ac.za/$\sim$zoaconf).

\section{Data}

The SDSS Data Release 1 consists of two slices of about 2.5 degrees
thick and $65-100$ degrees wide, centered on the celestial equator,
and of several regions at higher declinations.  In the present study we
have used only the equatorial slices.  From the general DR1 sample we
extracted the Northern and Southern slice samples using the following
criteria: the redshift interval $1000 \leq cz \leq 60000$~km s$^{-1}$,
the Petrosian $r^*$-magnitude interval $13.0 \leq r^* \leq 17.7$, the right
ascension and declination interval $145^{\circ} \leq RA \leq
250.0^{\circ}$ and $-1.25^{\circ} \leq DEC \leq 1.25^{\circ}$ for the
Northern slice, and $350^{\circ} \leq RA \leq 55.0^{\circ}$ and
$-1.25^{\circ} \leq DEC \leq 1.25^{\circ}$ for the Southern slice.
The number of galaxies extracted $N_{\rm gal}$ and the width $\Delta
RA$ are given in Table~\ref{Tab1}.

{\scriptsize
\begin{table*}
      \caption[]{Data on SDSS DR1 galaxies, clusters and superclusters}

         \label{Tab1}
      \[
         \begin{tabular}{ccccccccccc}
            \hline
            \noalign{\smallskip}
            Sample & $\Delta$RA &
      	$\alpha_E$&$M_E^{\ast}$&$\alpha_B$&$M_B^{\ast}$&
	$N_{\rm gal}$&$N_{\rm cl}$&$N_{\rm isol}$& $N_{\rm scl}$\\ 

            \noalign{\smallskip}
            \hline
            \noalign{\smallskip}

SDSS.N& $105^{\circ}$ &
$-1.06$&$-21.55$&$-1.05$&$-20.44$& 19783& 2754&10232&26 \\ 
SDSS.S& $66^{\circ}$&
$-1.06$&$-21.40$&$-1.05$&$-20.44$&11562&1451&6202&16\\ 
\\
            \noalign{\smallskip}
            \hline
         \end{tabular}
      \]
   \end{table*}
}

\begin{figure*}[ht]
\centering
\resizebox{0.45\textwidth}{!}{\includegraphics*{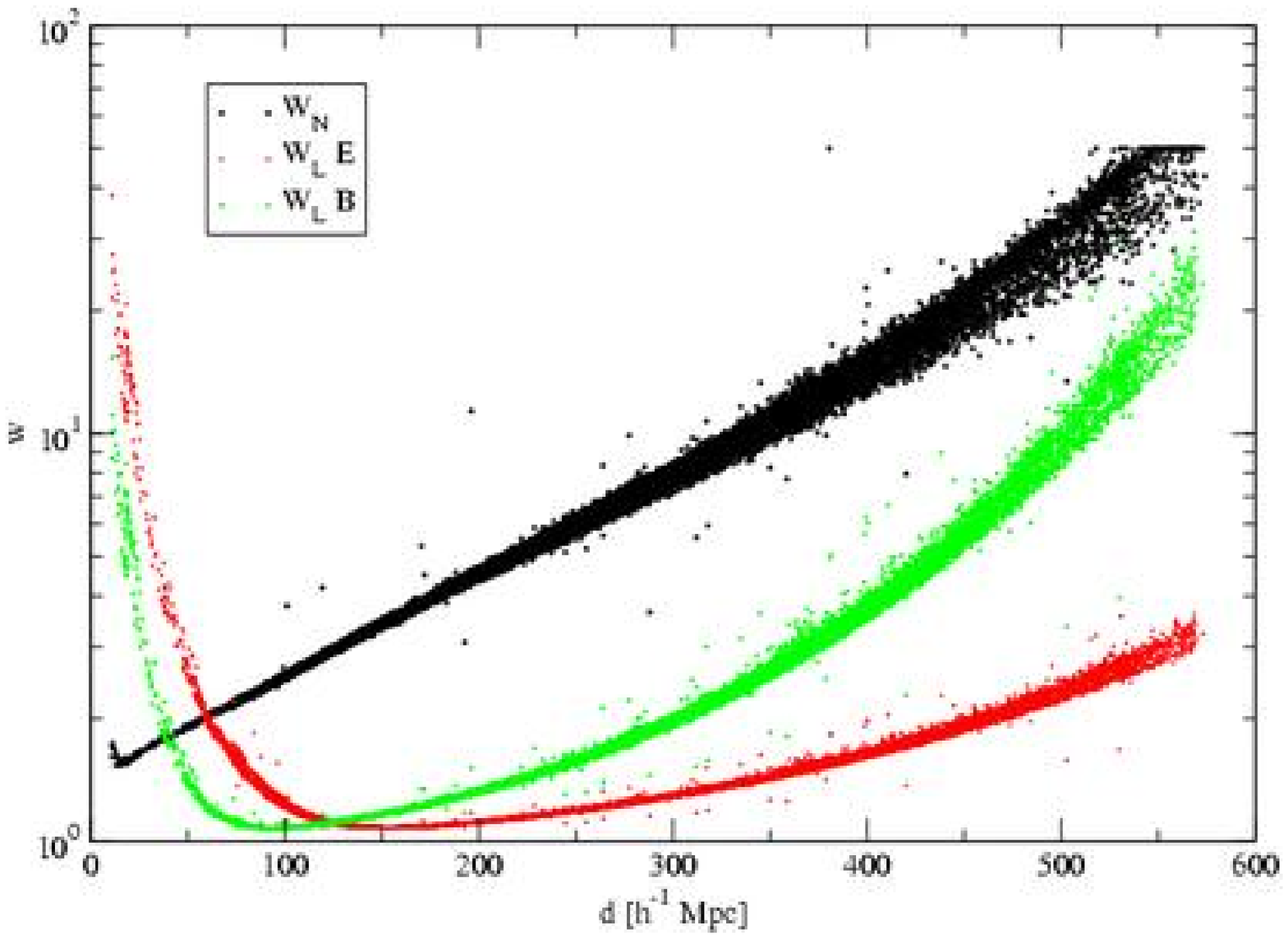}}
\hspace{2mm}
\resizebox{0.45\textwidth}{!}{\includegraphics*{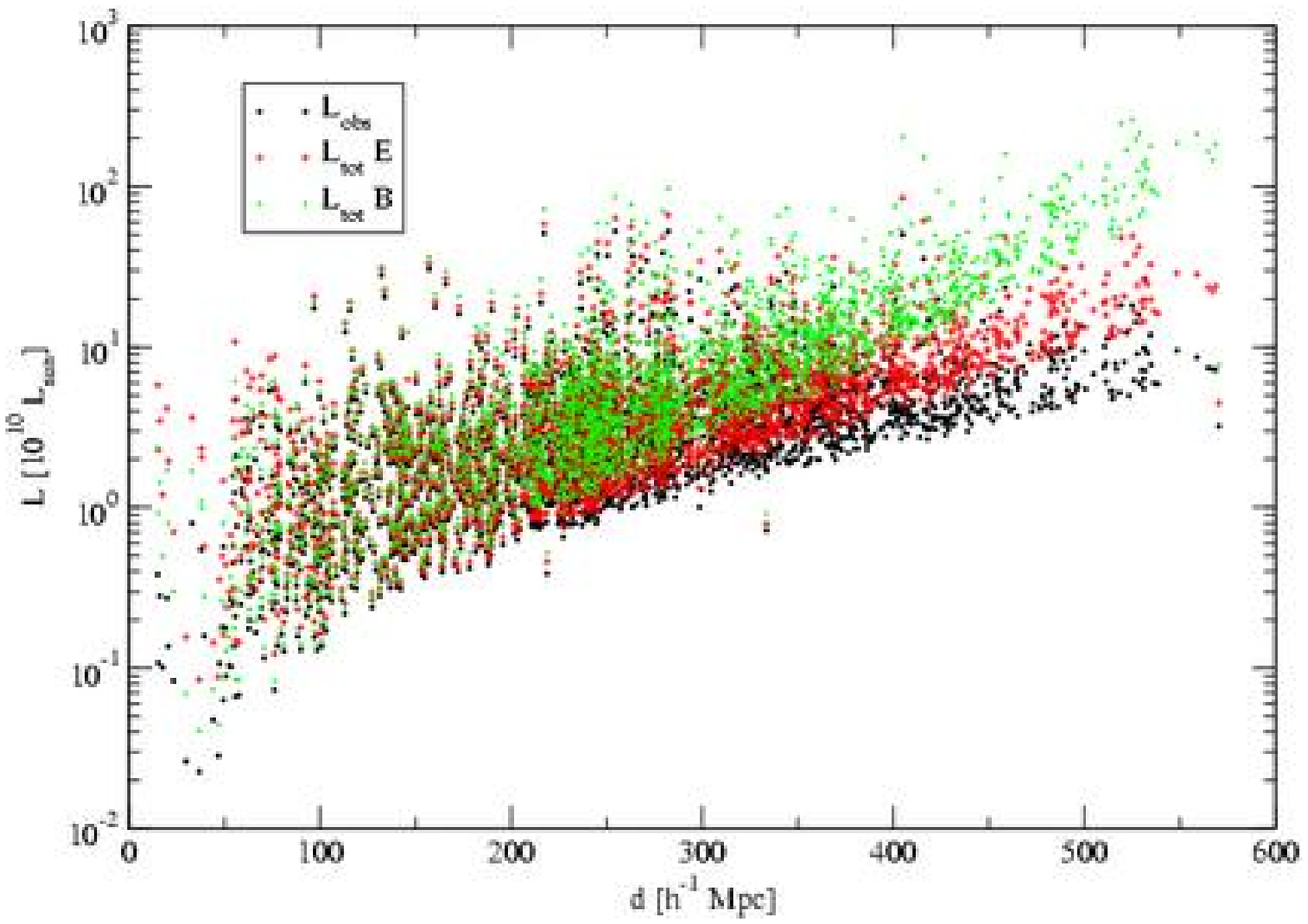}}
\hspace{2mm} 
\caption{The left panel shows the weights for observed galaxies,
  which are used to correct for
  invisible galaxies outside the observational window. Black symbols
  show the number-density weights, green and red symbols show the
  luminous-density weights, using the Blanton and Einasto sets of the Schechter
  function parameters, respectively. In the right panel we plot
  luminosities of galaxies: black symbols show observed luminosities,
  and green and red symbols show total luminosities for the Blanton and Einasto
  sets of the Schechter function parameters. 
}
\label{fig:1}
\end{figure*}

\begin{figure*}[ht]
\centering
\resizebox{0.45\textwidth}{!}{\includegraphics*{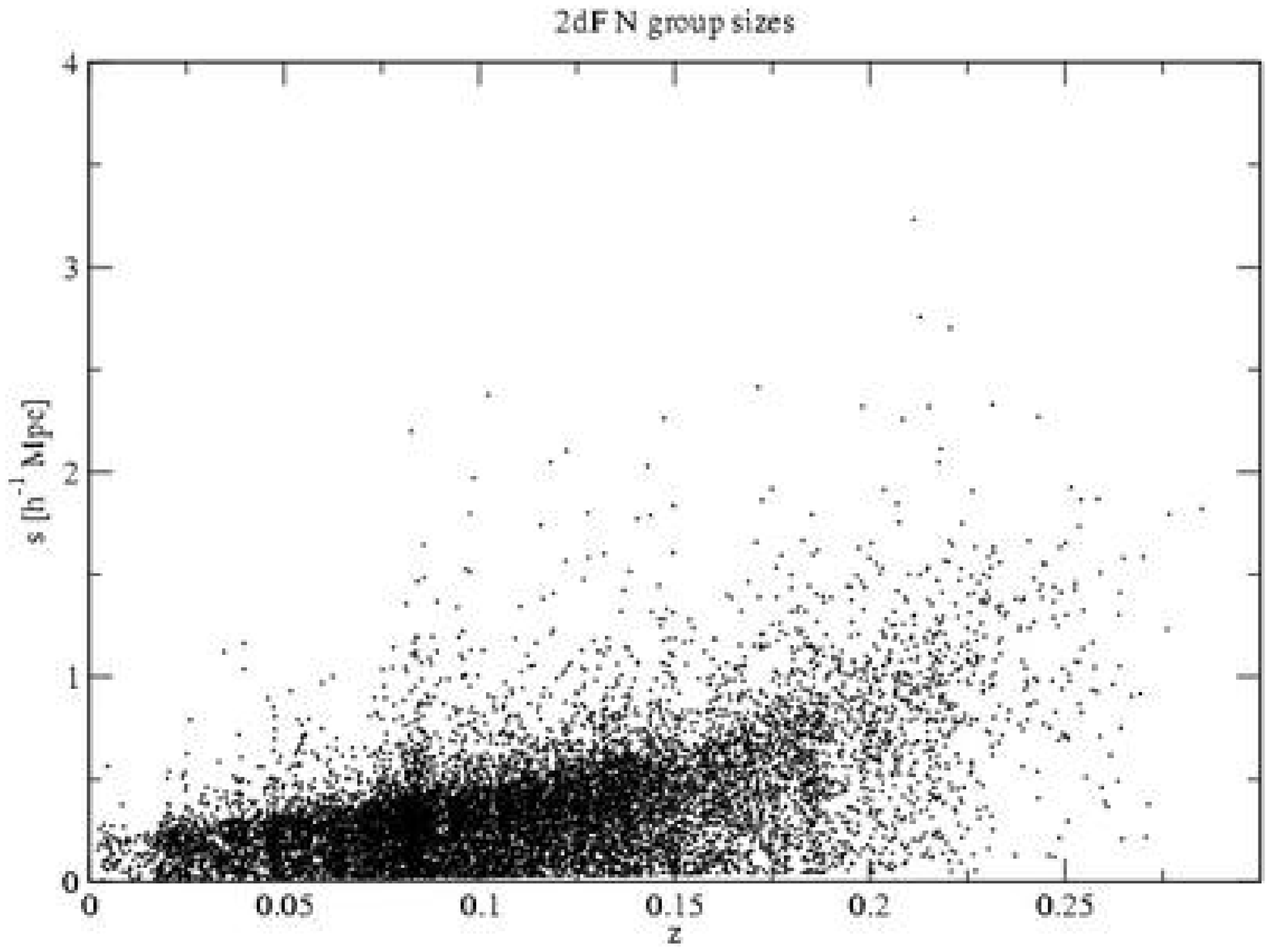}}\hspace{2mm}
\resizebox{0.45\textwidth}{!}{\includegraphics*{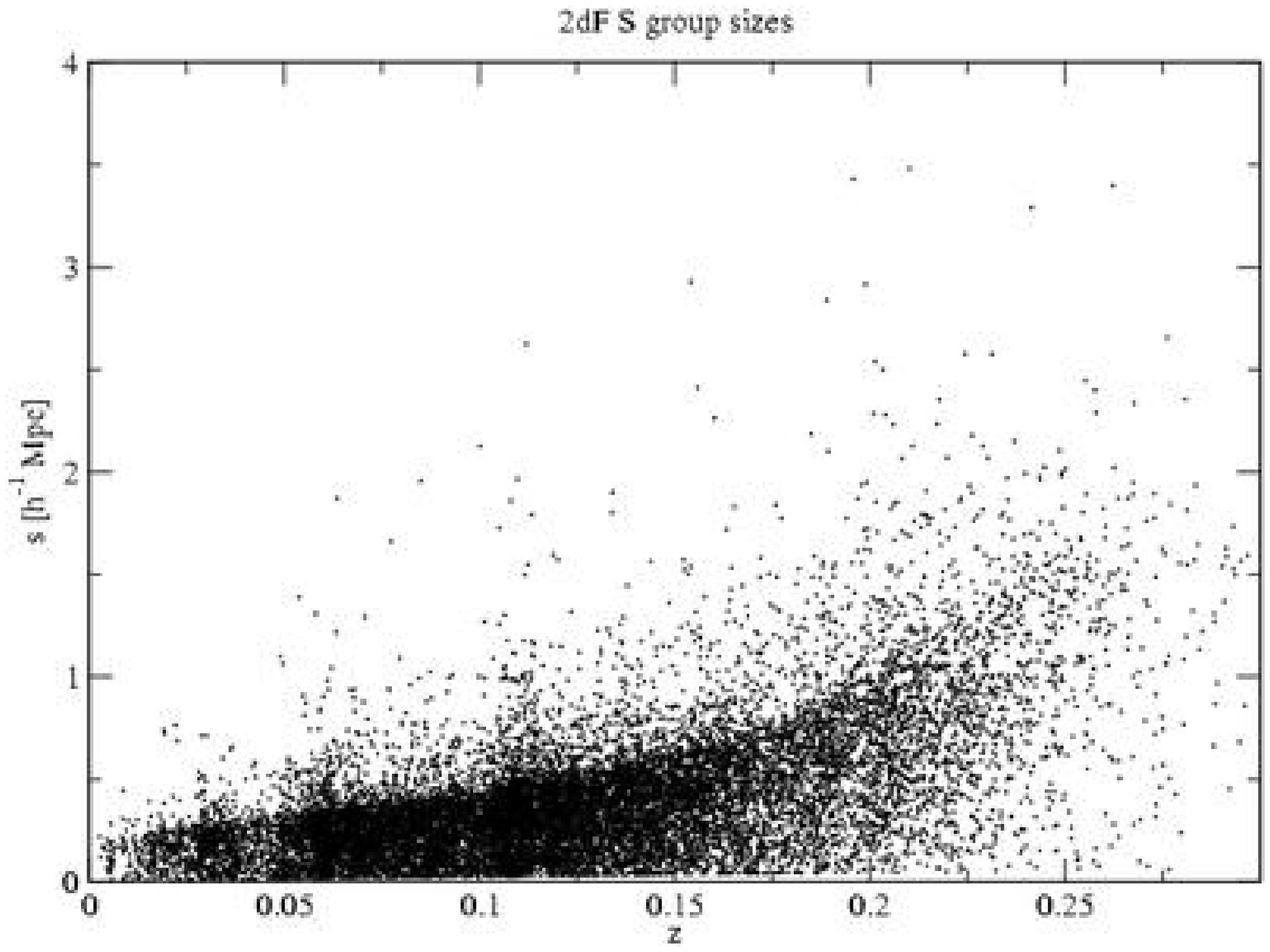}}\hspace{2mm}
\caption{Sizes of groups in the 2dF survey as found by Eke et
  al. \cite{eke04} as a function of their redshift $z$. The left panel shows
  the groups in the Northern Galactic hemisphere, the right panel shows
  the groups in the Southern hemisphere. 
}
\label{fig:2}
\end{figure*}

\begin{figure*}[ht]
\centering
\resizebox{0.50\textwidth}{!}
{\includegraphics*{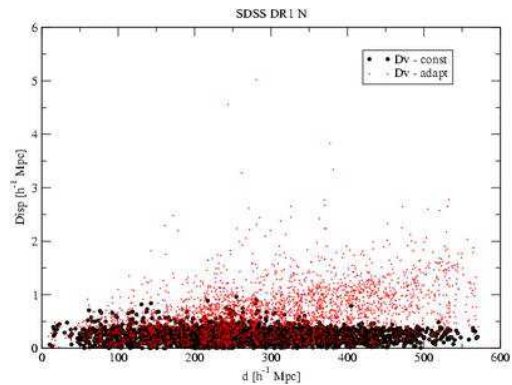}}
\hspace{2mm} 
\caption{ The  virial radii of
  groups(clusters) of galaxies in the Northern hemisphere, found with a 
  constant (black dots) and a variable (red crosses) search radius.
}
\label{fig:3}
\end{figure*}

\section{Data reduction}

Our data reduction procedure consists of several steps: (1)
calculation of the distance, the absolute magnitude, and the weight factor for
each galaxy of the sample; (2) finding groups/clusters of galaxies
using the friends-of-friends algorithm; (3) calculation of the density
field using an appropriate kernel and a selected smoothing length. 
When calculating
luminosities of galaxies, we regard every galaxy as a visible member of
a density enhancement (group or cluster) within the visible range of
absolute magnitudes, $M_1$ and $M_2$, corresponding 
to the observational window of apparent magnitudes
at the distance of the galaxy. This
assumption is based on observations of nearby galaxies, which indicate
that practically all galaxies belong to poor groups, like our own
Galaxy, where one bright galaxy is surrounded by a number of faint
satellites. Using this assumption we find halos, either halos of
single giant galaxies with their companions, or halos of
groups/clusters.  Further, we assume that the luminosity function
derived for a representative volume can be applied also for individual
groups and galaxies.

\begin{figure*}[ht]
\centering
\rotatebox{270}{
\resizebox{0.45\textwidth}{!}
{\includegraphics*{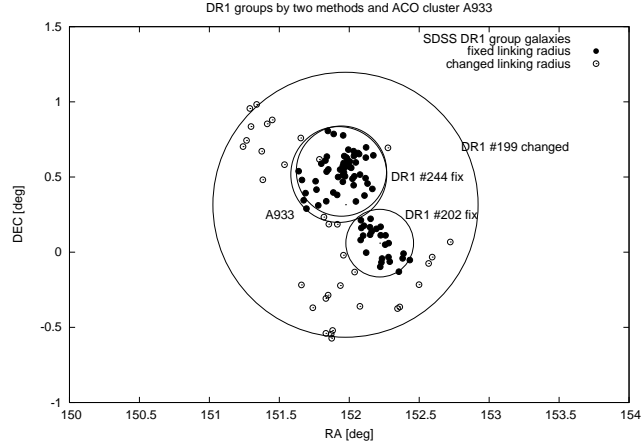}}}
\hspace{2mm}
\caption{An example of an observed complex of galaxies (in equatorial 
  coordinates),
  collected to a group using a variable search radius (open and filled
  circles), 
  and a constant search radius (filled circles only).  The group found
  with a variable search radius (DR1 199) consists of a cluster of galaxies,
  A933=DR1 244,  and of a group DR1 202, embedded in a cloud of loosely
  located galaxies.
}
\label{fig:4}
\end{figure*}

Calculation of distances, absolute magnitudes and weight
factors for galaxies has been described in detail in E03a.  In 
order to derive total luminosities of galaxies from their observed
luminosities, we applied the Schechter \cite{S76} function
with two sets of parameters. One set is based on the SDSS luminosity
function by Blanton et al. \cite{blanton}, the other set on the SDSS luminosity
function found in E03a, which yields better properties of clusters of
galaxies; the respective values of the characteristic luminosity
$M^{\ast}$ and the shape parameter $\alpha$ are given in
Table~\ref{Tab1}. The subscripts B and E denote the parameter sets by Blanton
et al. and Einasto et al., respectively. 

In Fig.~\ref{fig:1} we show the number-density and luminous-density
weights as a function of distance.  The number-density weight (shown here
for comparison only) increases very rapidly with distance, showing the
decrease of the number of galaxies in the observational window.
The luminous-density weights based on the Blanton parameter set are 
rather large at large distances, the weights based on the Einasto set
are much lower.  The right panel of
Fig.~\ref{fig:1} shows the observed and total luminosities of galaxies at
various distances.  The total luminosities are corrected to account for
galaxies outside of the observational window.

\begin{figure}[ht]
\centering
\resizebox{0.85\textwidth}{!}{\includegraphics*{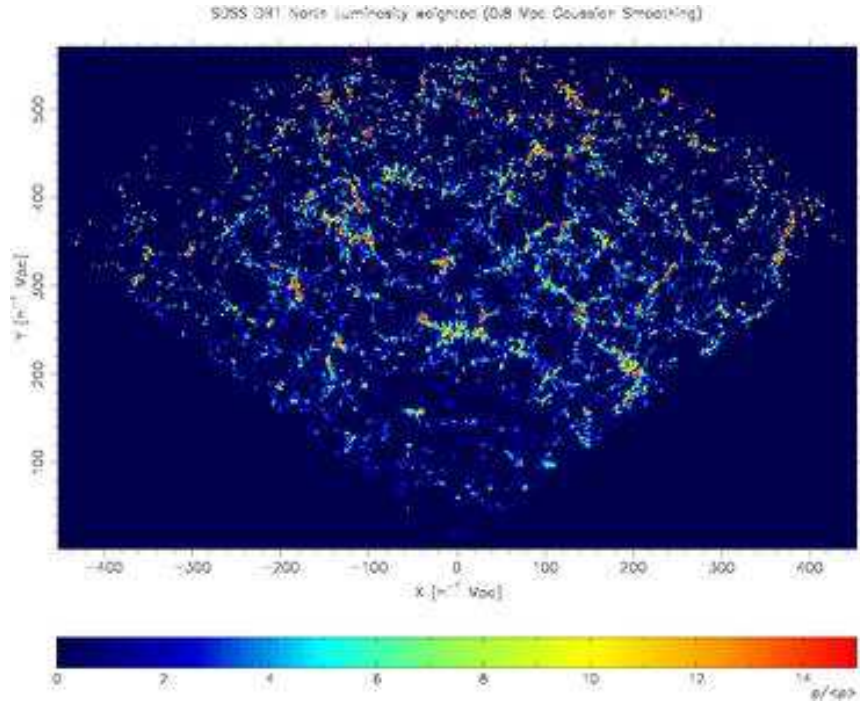}}
\hspace{2mm}
\caption{The high-resolution density field of the Northern slice of SDSS
  DR1. The density was calculated using Gaussian smoothing with 
  $\sigma=0.8$\Mpc. When calculating total luminosities of galaxies, the 
  Schechter function parameters of the set E were used.  }
\label{fig:5}
\end{figure}

The next step is the search for groups and clusters of galaxies.  Here
we used the conventional friends-of-friends algorithm by Zeldovich,
Einasto \& Shandarin \cite{zes82} and Huchra \& Geller \cite{hg82}
(hereafter ZES and HG, respectively).  These algorithms are
essentially identical with one difference: ZES used a constant search
radius to find neighbours whereas HG applied a variable search radius
depending on the volume density of galaxies at a particular distance
from the observer.  The variant with a variable search radius is widely
used, in particular by Eke et al. \cite{eke04}, for constructing the
group/cluster catalogue of 2dF redshift survey.  To see how good this
search algorithm is, we analyzed the mean radii of groups of this
catalogue. The results of the analysis are shown in Fig.~\ref{fig:2}.  We
see that the sizes of groups, generated with a variable search
radius, increase systematically with redshift $z$. Thus the population
of such groups is not homogeneous, the groups at the far-away side
of the sample are different from nearby groups.

For our group/cluster catalogue for SDSS DR1 we generated two versions, 
one with a constant search radius of
0.5~\Mpc, and the second with a variable search radius, which increased
with distance proportionally to the mean distance between galaxies
(as in HG). In both cases a radial linking length 500 km/s was used.
The results are shown in Fig.~\ref{fig:3}.  We see that
the mean virial radii of groups/clusters are practically constant for the
constant search radius case, and increase with distance for the
variable search radius case.  As an example, Fig.~\ref{fig:4} shows the
distribution of galaxies in the sky for one rich cluster.  Using a variable
search radius we include in the cluster two clusters of galaxies and a
number of nearby field galaxies.  One of these clusters is Abell 933;
a constant search radius collects to the cluster galaxies inside a
circle of radius 1.5~\Mpc, as used by Abell in his cluster
catalogue. The second cluster is less rich and is not included in the
Abell catalogue, its radius is also a bit smaller. 

In the following analysis we have used only the group/cluster
catalogue found with a constant search radius. The numbers of
groups/clusters found for both equatorial slices are given in
Table~\ref{Tab1}. 

\begin{figure*}[ht]
\centering
\resizebox{0.45\textwidth}{!}{\includegraphics*{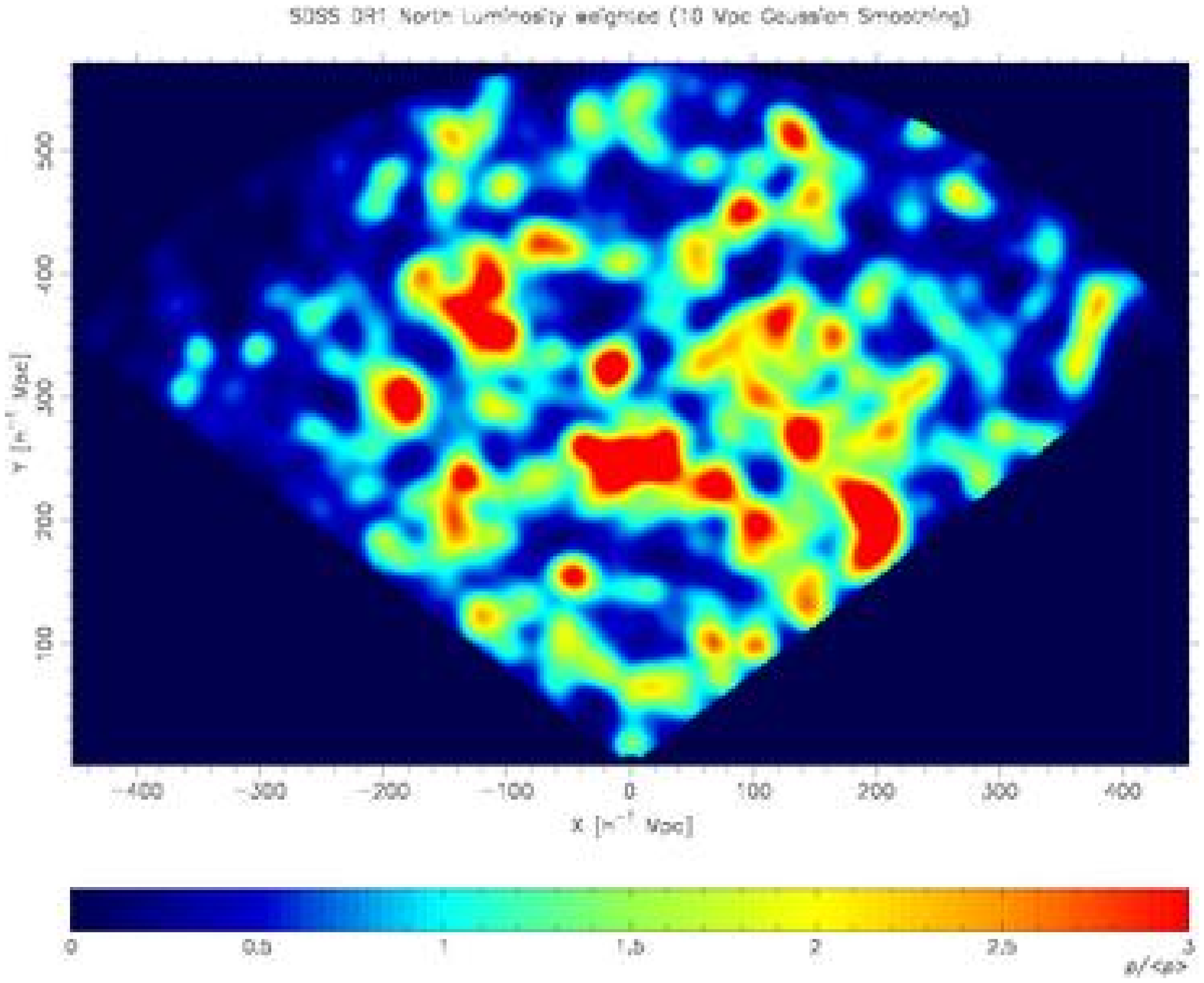}}
\hspace{2mm}
\resizebox{0.45\textwidth}{!}{\includegraphics*{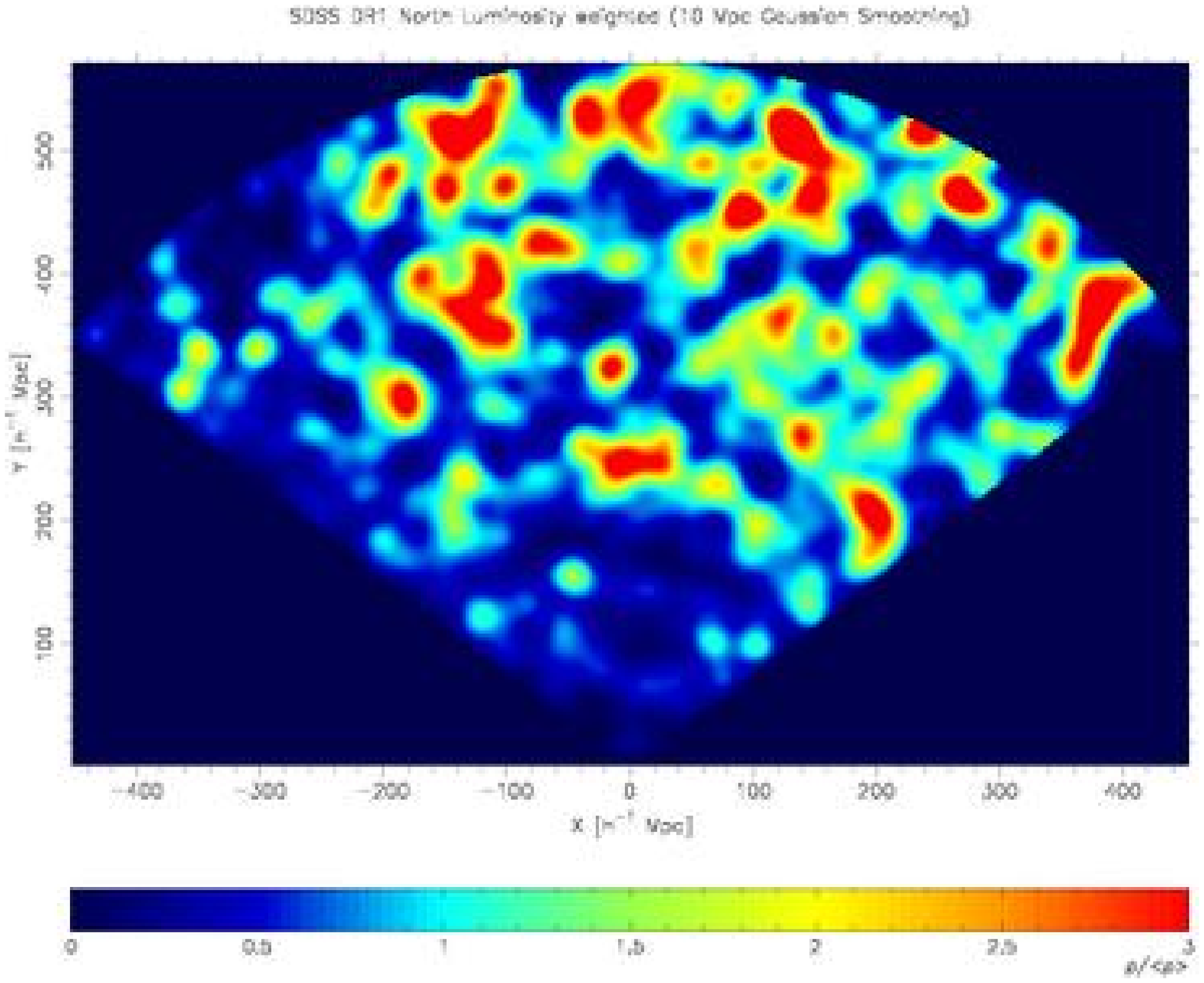}}
\hspace{2mm} 
\caption{The low-resolution luminosity density field of the Northern 
  slice of SDSS DR1 (in units of the mean density). The density was 
  calculated using Gaussian smoothing with 
  $\sigma=10$\Mpc. The left panel shows the luminosity density for the
  set E of the Schechter function parameters, and the right panel -- for
  the set B. }
\label{fig:6}
\end{figure*}

\section{Density field}

As both equatorial slices are very thin, we calculated only
2-dimensional luminosity density fields. As in E03a and E03b, we
calculated the high-resolution density field using Gaussian smoothing
with $\sigma=0.8$\Mpc, and the low-resolution field with
$\sigma=10$\Mpc.  The high-resolution field was found using the
Schechter parameters of set E, and the low-resolution field with both
parameter sets.  The results are shown in Figs.~\ref{fig:5} and
\ref{fig:6}, respectively. The low-resolution field was used to define
superclusters as connected over-density regions. As in E03a, we used
the density thresholds 1.8--2.1 to find superclusters.  At lower
thresholds superclusters start to merge into percolating systems,
violating the definition of superclusters as largest but still
isolated high-density regions.  For higher thresholds the number of
superclusters rapidly decreases (many of them have lower peak
density).

The comparison of high- and low-resolution fields yields information
on the fine structure of superclusters of various size and luminosity.
We see that within superclusters clusters may form a single filament,
a branching system of filaments, or a more or less diffuse cloud of
clusters. Also we see that clusters themselves have various richness:
in massive superclusters most clusters are very bright, in poor
superclusters galaxy systems are also poor.

Fig.~\ref{fig:6} shows that the mean luminosity of superclusters found
for the set B of the Schechter parameters increases considerably with
distance. In other words, this parameter set gives too high weights for
galaxies outside the visibility window.  In contrast, the parameter
set E yields superclusters that are a bit too luminous at medium
distances from the observer.  We note that the superclusters found by E03a
and E03b for SDSS EDR and LCRS have luminosities which are, in the
mean, independent of the distance from the observer.  This shows that
a very careful choice of the parameters of the Schechter function is
essential.  In the following analysis we used the parameter set E to
find environmental densities for clusters of galaxies, as in this case
the distance dependence of supercluster luminosities is much less than
for the set B.

\begin{figure*}[ht]
\centering
\resizebox{0.45\textwidth}{!}{\includegraphics*{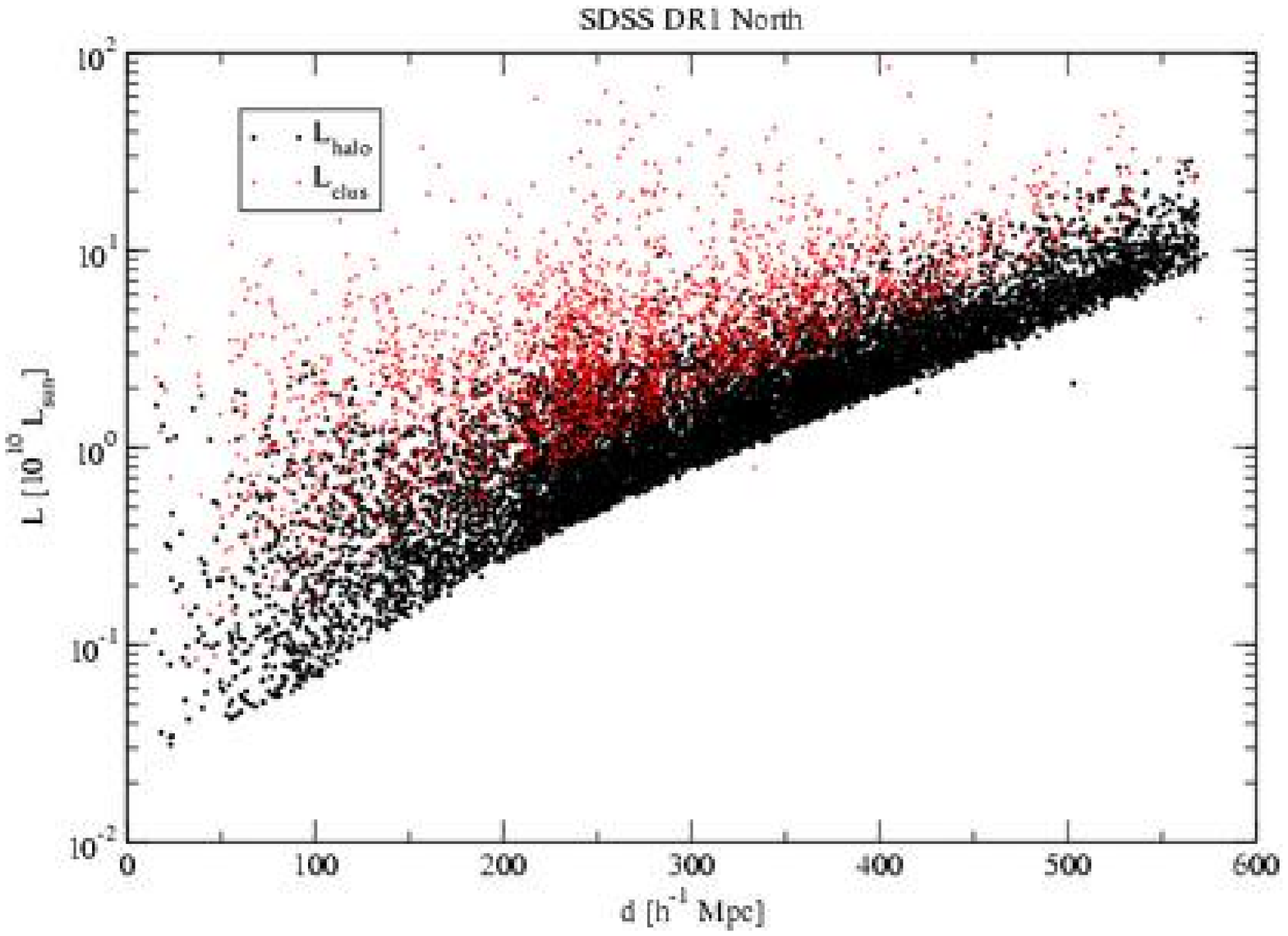}}
\hspace{2mm}
\resizebox{0.45\textwidth}{!}{\includegraphics*{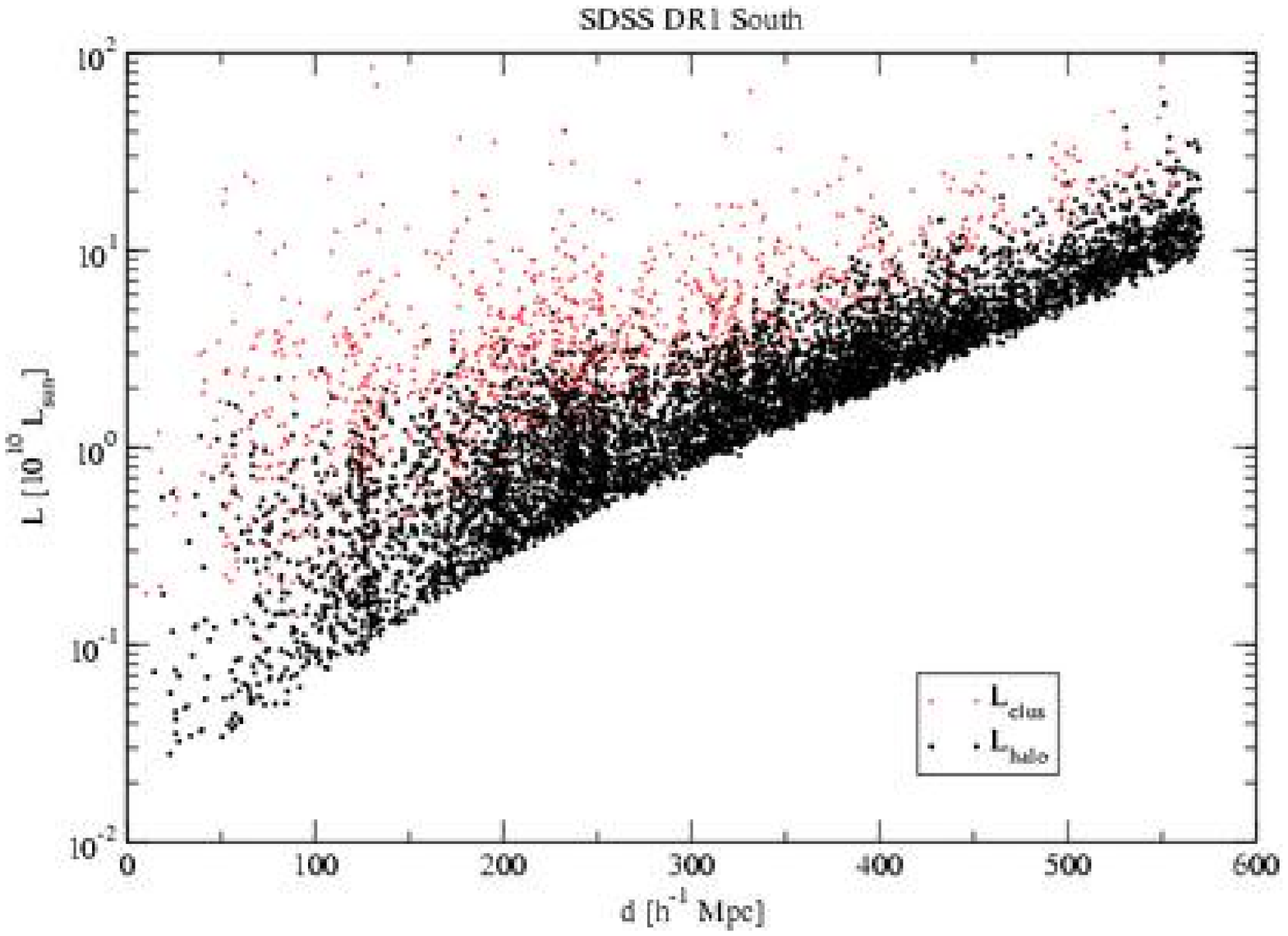}}
\hspace{2mm} 
\caption{Luminosities of groups/clusters at different distances,
  corrected for galaxies outside the visibility window. Red symbols
  denote groups with at least two visible galaxies, black symbols denote
  halos containing only one galaxy in the visibility window. The left 
  panel shows the results for the SDSS Northern slice, the right panel --
  for the Southern slice.}
\label{fig:7}
\end{figure*}

\begin{figure*}[ht]
\centering
\resizebox{0.50\textwidth}{!}{\includegraphics*{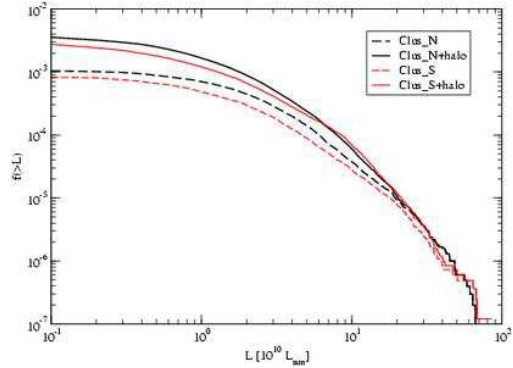}}
\hspace{2mm}
\caption{Luminosity functions for SDSS DR1 groups/clusters. Dashed
  lines show functions found using clusters with at least two
  galaxies in the observational window, and solid lines show the
  luminosity functions for all
  groups/clusters, including halos with only one galaxy in the observational
  window. The set E of the Schechter
  function parameters was used to calculate total luminosities.  }
\label{fig:8}
\end{figure*}

\section{Properties of clusters and superclusters}

\begin{figure*}[ht]
\centering
\resizebox{0.45\textwidth}{!}{\includegraphics*{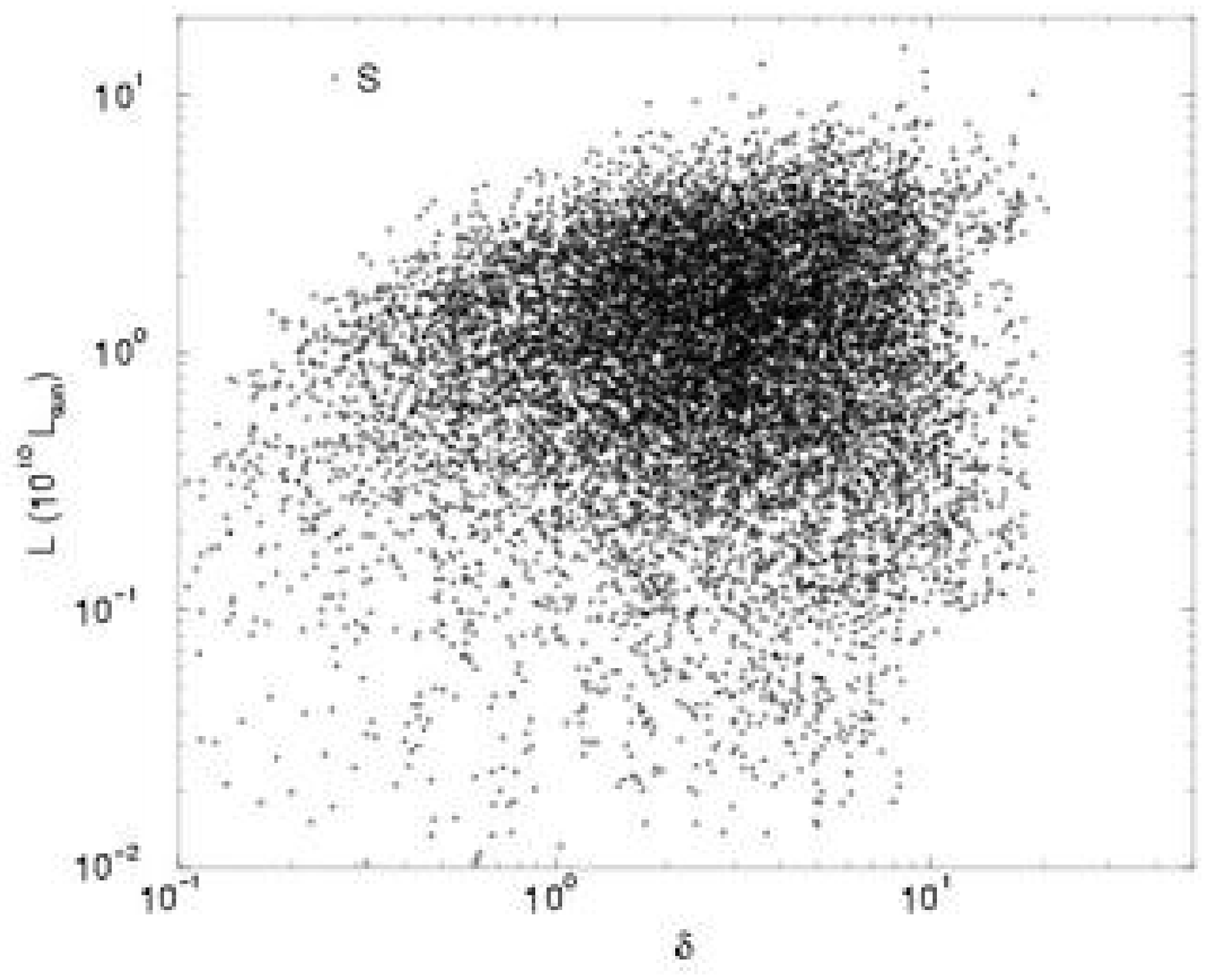}}
\resizebox{0.45\textwidth}{!}{\includegraphics*{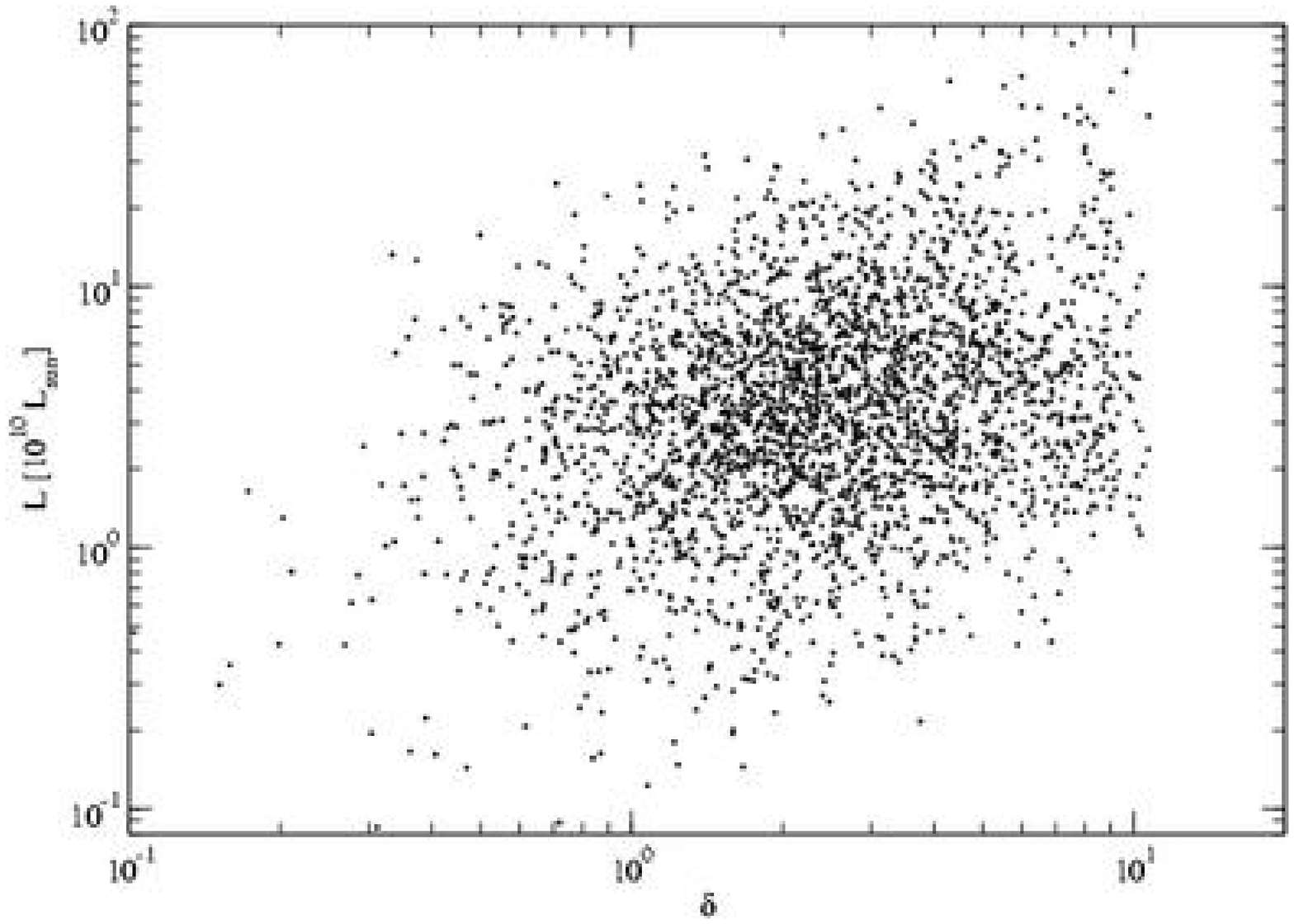}}
\hspace{2mm}
\caption{The left panel shows the luminosities of galaxies in various
  density environments for the 
  SDSS EDR Southern slice. The environmental density was calculated using
  Gaussian smoothing with $\sigma=2$\Mpc.  The right panel  shows the
  luminosities of the SDSS DR1 Northern slice 
  clusters as a function of the environmental density, found with
  Gaussian smoothing of the luminous density field with $\sigma=10$\Mpc. 
}
\label{fig:9}
\end{figure*}

Fig.~\ref{fig:7} shows luminosities of groups/clusters at different
distances from the observer.  We see that there exists a well-defined
lower limit of cluster luminosities at larger distances; this limit is
linear in the $\log L - d$ plot.  Such behaviour is expected, as at
large distances an increasing fraction of clusters does not contain
any galaxies bright enough to fall into the observational window of
absolute magnitudes, $M_1 \dots M_2$.  The limit is lower for groups
containing only one galaxy in the visibility window; these groups are
actually halos with one bright galaxy surrounded by faint
companions. The low-luminosity limit for halos and
groups containing at least two galaxies in the visibility window is two
times higher, as expected (this factor corresponds to the case when both
galaxies in the visibility window have equal luminosities). 

Fig.~\ref{fig:8} shows the integrated luminosity function of
groups/clusters for the SDSS DR1 Northern and Southern samples.  The
absence of low-luminosity clusters at large distances has been taken
into account by a standard weighting procedure (for details see
E03a).  The luminosity function was calculated separately for
groups/clusters with at least two visible galaxies, and for all
groups/clusters including halos with only one visible galaxy in the
visibility window.  In both cases the numbers of
clusters have been corrected to take into account selection effects.
Our calculations show that in the second case the
number of groups/clusters per unit volume is larger for the low luminosity
range of the luminosity function.  
This difference is probably due to the small number of low-luminosity
groups. 

The volume density of groups/clusters according to the SDSS DR1 data is $3
\times 10^{-3}$ (\Mpc)$^{-3}$ for $L \geq 10^9$ $L_{\odot}$ groups/clusters.
This estimate is in fairly good agreement with the estimates of the number
density of groups based on the group mass function by Girardi \& Giuricin
\cite{gg00}.

Let us now consider properties of galaxies and clusters in various
environments.  We shall use the density found with 2~\Mpc\
smoothing as an environmental parameter to describe the surrounding
density of galaxies.  Similarly, we use the density found with 
10~\Mpc\ smoothing as the global density in the supercluster
environment of clusters.  The luminosity of galaxies as a function of
the environmental density is shown in the left panel of
Fig.~\ref{fig:9}.  This Figure 
demonstrates the well-known density dependence of galaxies in systems:
in groups and clusters the centrally located main galaxy has
considerably higher luminosity than the surrounding galaxies.

\begin{figure*}[ht]
\centering
\resizebox{0.45\textwidth}{!}{\includegraphics*{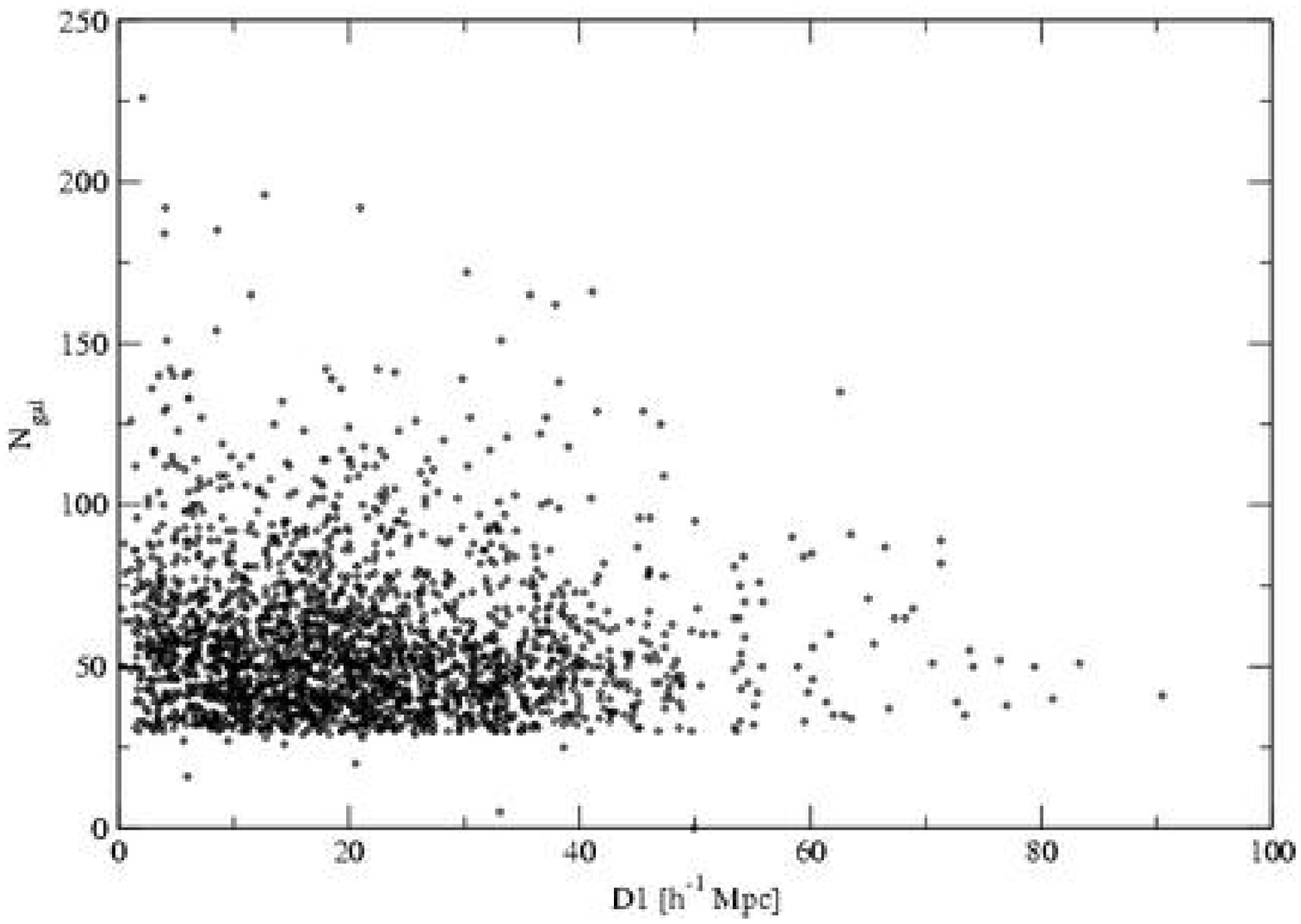}}
\hspace{2mm}
\resizebox{0.45\textwidth}{!}{\includegraphics*{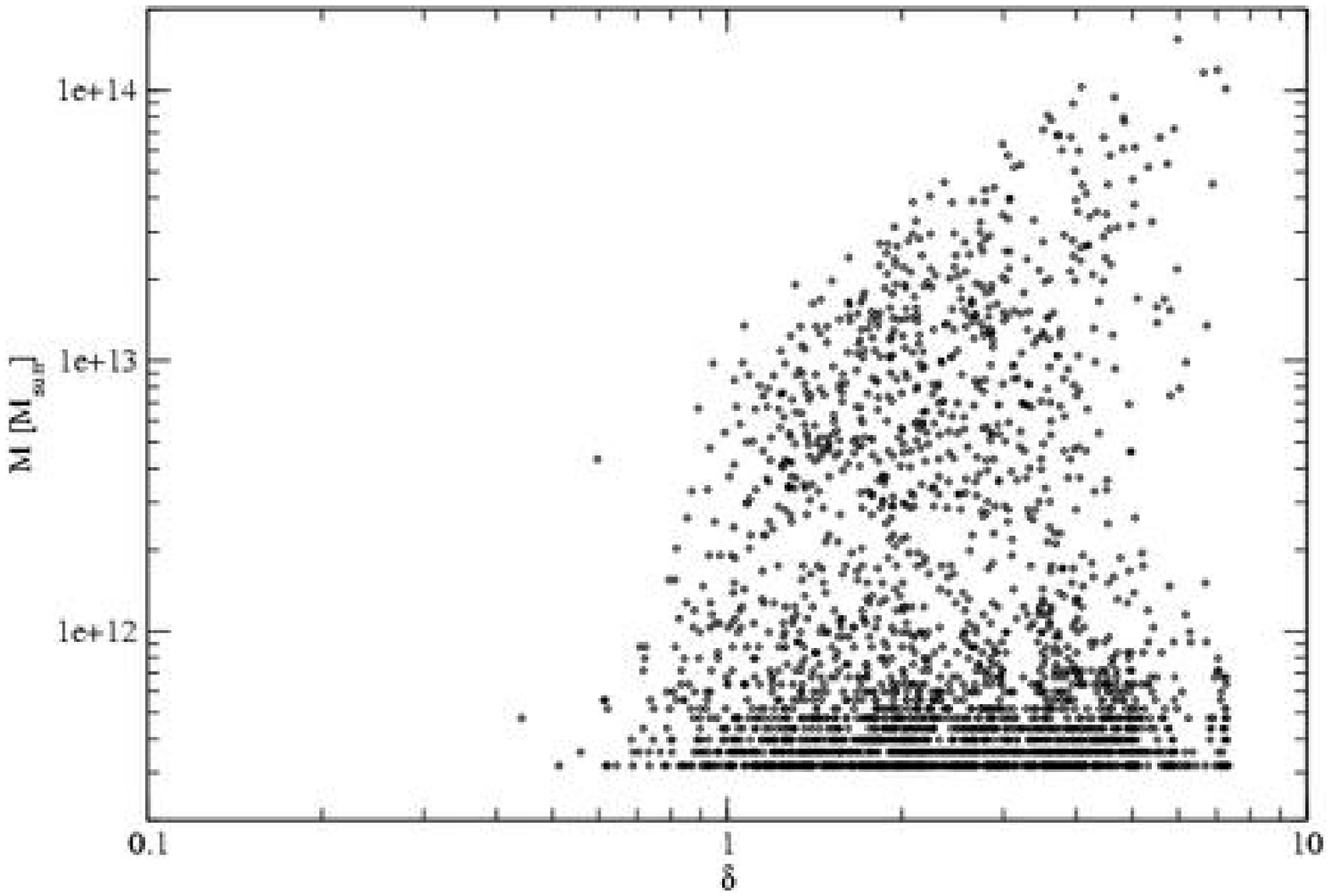}}
\hspace{2mm}
\caption{The left panel shows Abell cluster richness (the number of
  galaxies in clusters) as a function of the distance to the 1st nearest
  neighbour, used as the environmental density parameter.  The right
  panel shows  masses of 
  clusters in environments of various density in the N-body simulation. 
 The density was found, using an Epanechnikov kernel with the radius 10~\Mpc.
}
\label{fig:10}
\end{figure*}

On supercluster scales this effect is seen on the right panel of
Fig.~\ref{fig:9}.  There 
is a clear correlation between the luminosity of DF-clusters and their
environmental density.  Luminous clusters are predominantly located in
high-density regions, and low-luminosity clusters -- in low-density
regions.  This tendency can be seen also in Fig.~\ref{fig:5}.
Here densities are color-coded, and we see that small clusters in voids
have blue color, which indicates medium and small densities, whereas
rich clusters having red color populate dominantly the central
high-density regions of superclusters.  For Abell clusters the
dependence of the cluster richness on the density of the environment is
shown on left panel of Fig.~\ref{fig:10}.  The domination of faint
galaxies in void regions was noticed by Lindner et al. \cite{l95}.
The environmental enhancement 
effect was found in the vicinity of rich clusters of galaxies by E03c and
E03d.

\section{Comparison with N-body models}

The final step in our study is comparison of observational data
with numerical simulations.  We have used in this preliminary stage of
the study a simulation with $128^3$ particles in a 100~\Mpc\ cube.
Conventional cosmological parameters were used: the matter density
$\Omega_m = 0.3$,  the dark energy density $\Omega_{\Lambda} = 0.7$,
the power spectrum amplitude parameter $\sigma_8 = 0.8$.  
Clusters were identified by the
FoF algorithm with the search radius parameter $b=0.2$.   

The mass of groups/clusters is plotted in the Fig.~\ref{fig:10} as a
function of the density of the environment, calculated with the
Epanechnikov kernel 10~\Mpc.  Here the dependence of the cluster mass
on the density of the environment is very well expressed: most massive
clusters in high-density environments have masses that are about two
orders of magnitude higher than the masses of most massive clusters in
low-density environments.

\begin{figure*}[ht]
\centering
\resizebox{0.55\textwidth}{!}{\includegraphics*{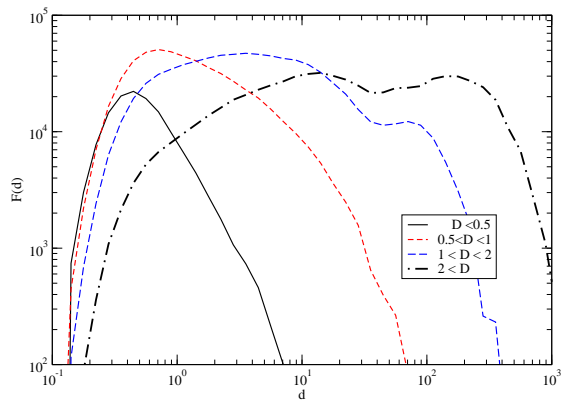}}
\hspace{2mm}
\caption{The distribution of particles as a function of the local density
  $d$ of the environment in a N-body model. The local density of particles
  was found by interpolation of the unsmoothed density field.  The
  distribution is shown for various regions of the global density $D$.
}
\label{fig:11}
\end{figure*}

To understand better the dependence of cluster properties on the
environment we studied the distribution of particle densities. We
attributed to
every particle in the simulation two density values,
corresponding to the density at the location of the particle,
found without smoothing and with smoothing.  In the second case we
used smoothing with the Epanechnikov kernel of a smoothing radius
10~\Mpc.  This second density was used as a global environmental
parameter. The whole simulation box was divided into 4 regions
according to the global density $D$ (in units of the mean density): 
$D \geq 2$, $1 \leq D < 2$, $0.5
\leq D < 1$, and $D < 0.5$. These global density regions correspond
approximately to superclusters, rich and poor filaments, and systems
in large voids. The results are shown in Fig.~\ref{fig:11}.  We see that
in superclusters (regions of high global density) the majority of
particles are located in systems of high local density (rich clusters
of galaxies). Further we see that a small fraction of particles is
located in poor clusters or groups, and even a smaller fraction of
particles form the void population. These
particles have the local density less then 1 and cannot cluster, since
galaxy formation starts only in the case if the local density exceeds
a certain threshold, much higher than the mean density (see Press \&
Schechter \cite{ps74}).  With a decreasing global density the fraction
of particles in very rich clusters decreases, most particles belong to
intermediate rich groups and clusters, and the fraction of particles in
voids increases.  Finally, in the void regions of the lowest global density
most particles have local density less than 1, and a small fraction of
particles are clustered forming very poor groups of galaxies (with local
densities less than 10). 

Differences in the evolution of high and low-density regions were
studied by Frisch et al. \cite{f95}, they are
evident also in movies prepared by Gottl\"ober and M\"uller in the Potsdam
Astrophysical Institute.  One movie was made for a high-density region
(the central cluster of a rich supercluster), the other movie for a large
under-dense region of 20~\Mpc\ diameter (in co-moving coordinates).  In
the high-density region the galaxy formation starts at an early epoch
and leads to a rapid merging of numerous small clumps to a very rich
cluster.  In the large void region a filamentary network of clustered
particles also forms, but these filaments are relatively poor and
contain only a small number of knots which can be identified with
dwarf galaxies and very poor groups.

Finally we used numerical simulations to investigate
properties of superclusters.  In addition to the density fields derived
with two different smoothing lengths we calculated also the
gravitational potential field with and without smoothing.  In the
unsmoothed potential field all clusters of galaxies are seen as local
attractors, which distort the otherwise smooth potential field.  Rich
superclusters can be identified as large depressions in the potential
field.  However, in contrast to the density field derived with a large
smoothing length, it was impossible to define low-mass superclusters
using the potential field.  The reason is simple -- the potential
field is much shallower than the density field, and low-mass
superclusters do not generate a depression in the potential field deep
enough to make the identification of the supercluster possible. 

\begin{figure*}[ht]
\centering
\resizebox{0.55\textwidth}{!}{\includegraphics*{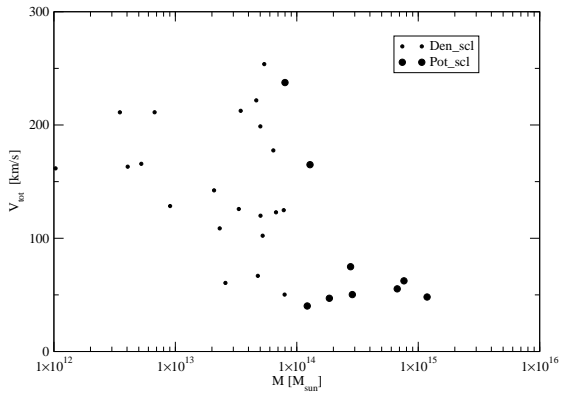}}
\hspace{2mm}
\caption{The mean total velocity of superclusters in a N-body simulation
  as a function of their mass (in Solar mass units).  Large symbols
  denote for superclusters which can be identified both in the
  gravitational potential as well as in the smoothed density field, small
  symbols denote superclusters which were found only in the density field. }
\label{fig:12}
\end{figure*}

To find superclusters in a simulation we applied the same procedure as
for real superclusters, i.e. they were identified as relatively
isolated high-density regions in the low-resolution density
field. Experimentation with various threshold density levels showed that
the optimal density level to extract superclusters lies in the interval $1.8
\dots 2.1$ (in units of the mean density). Similar threshold densities
were also applied to find real superclusters in both the SDSS and LCRS
galaxy samples by E03a and E03b.  We calculated masses of
superclusters by adding 
masses of clusters within supercluster boundaries.  Next we calculated
mean velocities of superclusters by summing up velocities of all
clusters in superclusters.  These velocities as a function of supercluster
mass are shown in Fig.~\ref{fig:12}.  In this figure, massive superclusters 
which can
be identified both in the density, as well as in the potential field, are
plotted by large symbols.  We see that these massive superclusters have low
bulk velocities, and there is a rather sharp transition to less
massive superclusters, which have much larger bulk velocities.  In
other words, massive superclusters can be considered great attractors,
whereas low-mass superclusters are much smaller attractors.  Presently
this phenomenon has been established only in a relatively small
simulation box.  

To conclude we can say that the study of clusters in the SDSS DR1 using
3-dimensional information has confirmed our preliminary results based
on the SDSS EDR and LCRS on the environmental dependence of cluster
properties.  Numerical simulations show that in large underdense
regions most particles form a rarefied population of pregalactic
matter whereas in large overdense regions most particles form a
clustered population in rich clusters. 
 Comparison of observational data with results of numerical
simulation shows that superclusters can be divided into two classes;
very massive superclusters are great attractors, low-mass
superclusters are  small attractors having  larger bulk
velocities.

\end{document}